\documentclass[useAMS,usenatbib,usegraphicx]{mn2e}

\title[LBVs as Supernova Progenitors]{On Luminous Blue Variables as
  the Progenitors of Core-Collapse Supernovae, especially Type IIn
  Supernovae} \author[Dwarkadas]{V. V. Dwarkadas,$^{1}$\thanks{E-mail:
    vikram@oddjob.uchicago.edu}\\$^{1}$Department of Astronomy and
  Astrophysics, U Chicago, 5640 S Ellis Ave, Chicago, IL 60637}
  
\begin{document}
\newcommand{\vper}{\mbox{${v_{\perp}}$}}
\newcommand{\vpar}{\mbox{${v_{\parallel}}$}}
\newcommand{\uper}{\mbox{${u_{\perp}}$}}
\newcommand{\vperout}{\mbox{${{v_{\perp}}_{o}}$}}
\newcommand{\uperout}{\mbox{${{u_{\perp}}_{o}}$}}
\newcommand{\vperin}{\mbox{${{v_{\perp}}_{i}}$}}
\newcommand{\uperin}{\mbox{${{u_{\perp}}_{i}}$}}
\newcommand{\upar}{\mbox{${u_{\parallel}}$}}
\newcommand{\uparout}{\mbox{${{u_{\parallel}}_{o}}$}}
\newcommand{\vparout}{\mbox{${{v_{\parallel}}_{o}}$}}
\newcommand{\uparin}{\mbox{${{u_{\parallel}}_{i}}$}}
\newcommand{\vparin}{\mbox{${{v_{\parallel}}_{i}}$}}
\newcommand{\dout}{\mbox{${\rho}_{o}$}}
\newcommand{\din}{\mbox{${\rho}_{i}$}}
\newcommand{\da}{\mbox{${\rho}_{1}$}}
\newcommand{\mfast}{\mbox{$\dot{M}_{f}$}}
\newcommand{\mslow}{\mbox{$\dot{M}_{a}$}}
\newcommand{\beqn}{\begin{eqnarray}}
\newcommand{\eeqn}{\end{eqnarray}}
\newcommand{\be}{\begin{equation}}
\newcommand{\ee}{\end{equation}}
\newcommand{\noi}{\noindent}
\newcommand{\ftheta}{\mbox{$f(\theta)$}}
\newcommand{\gtheta}{\mbox{$g(\theta)$}}
\newcommand{\ltheta}{\mbox{$L(\theta)$}}
\newcommand{\stheta}{\mbox{$S(\theta)$}}
\newcommand{\utheta}{\mbox{$U(\theta)$}}
\newcommand{\xitheta}{\mbox{$\xi(\theta)$}}
\newcommand{\vs}{\mbox{${v_{s}}$}}
\newcommand{\ro}{\mbox{${R_{0}}$}}
\newcommand{\pa}{\mbox{${P_{1}}$}}
\newcommand{\va}{\mbox{${v_{a}}$}}
\newcommand{\vo}{\mbox{${v_{o}}$}}
\newcommand{\vp}{\mbox{${v_{p}}$}}
\newcommand{\vw}{\mbox{${v_{w}}$}}
\newcommand{\vf}{\mbox{${v_{f}}$}}
\newcommand{\lprime}{\mbox{${L^{\prime}}$}}
\newcommand{\uprime}{\mbox{${U^{\prime}}$}}
\newcommand{\sprime}{\mbox{${S^{\prime}}$}}
\newcommand{\xiprime}{\mbox{${{\xi}^{\prime}}$}}
\newcommand{\mdot}{\mbox{$\dot{M}$}}
\newcommand{\msun}{\mbox{$M_{\odot}$}}
\newcommand{\yr}{\mbox{${\rm yr}^{-1}$}}
\newcommand{\kms}{\mbox{${\rm km} \;{\rm s}^{-1}$}}
\newcommand{\lambdav}{\mbox{${\lambda}_{v}$}}
\newcommand{\lequ}{\mbox{${L_{eq}}$}}
\newcommand{\eqpratio}{\mbox{${R_{eq}/R_{p}}$}}
\newcommand{\ra}{\mbox{${r_{o}}$}}
\newcommand{\bfig}{\begin{figure}[h]}
\newcommand{\efig}{\end{figure}}
\newcommand{\tone}{\mbox{${t_{1}}$}}
\newcommand{\done}{\mbox{${{\rho}_{1}}$}}
%%%%%%%%
\newcommand{\dsn}{\mbox{${\rho}_{SN}$}}
\newcommand{\dzero}{\mbox{${\rho}_{0}$}}
\newcommand{\ve}{\mbox{${v}_{e}$}}
\newcommand{\vej}{\mbox{${v}_{ej}$}}
\newcommand{\Mch}{\mbox{${M}_{ch}$}}
\newcommand{\mej}{\mbox{${M}_{e}$}}
\newcommand{\Mst}{\mbox{${M}_{ST}$}}
\newcommand{\dam}{\mbox{${\rho}_{am}$}}
\newcommand{\Rst}{\mbox{${R}_{ST}$}}
\newcommand{\Vst}{\mbox{${V}_{ST}$}}
\newcommand{\Tst}{\mbox{${T}_{ST}$}}
\newcommand{\no}{\mbox{${n}_{0}$}}
\newcommand{\Efif}{\mbox{${E}_{51}$}}
\newcommand{\rsh}{\mbox{${R}_{sh}$}}
\newcommand{\msh}{\mbox{${M}_{sh}$}}
\newcommand{\vsh}{\mbox{${V}_{sh}$}}
\newcommand{\vrev}{\mbox{${v}_{rev}$}}
\newcommand{\rpr}{\mbox{${R}^{\prime}$}}
\newcommand{\mpr}{\mbox{${M}^{\prime}$}}
\newcommand{\vpr}{\mbox{${V}^{\prime}$}}
\newcommand{\tpr}{\mbox{${t}^{\prime}$}}
\newcommand{\cone}{\mbox{${c}_{1}$}}
\newcommand{\ctwo}{\mbox{${c}_{2}$}}
\newcommand{\cthree}{\mbox{${c}_{3}$}}
\newcommand{\cfour}{\mbox{${c}_{4}$}}
\newcommand{\Te}{\mbox{${T}_{e}$}}
\newcommand{\Ti}{\mbox{${T}_{i}$}}
\newcommand{\Ha}{\mbox{${H}_{\alpha}$}}
\newcommand{\Rprime}{\mbox{${R}^{\prime}$}}
\newcommand{\Vprime}{\mbox{${V}^{\prime}$}}
\newcommand{\Tprime}{\mbox{${T}^{\prime}$}}
\newcommand{\Mprime}{\mbox{${M}^{\prime}$}}
\newcommand{\rprime}{\mbox{${r}^{\prime}$}}
\newcommand{\rfprime}{\mbox{${r}_f^{\prime}$}}
\newcommand{\vprime}{\mbox{${v}^{\prime}$}}
\newcommand{\tprime}{\mbox{${t}^{\prime}$}}
\newcommand{\mprime}{\mbox{${m}^{\prime}$}}
\newcommand{\Me}{\mbox{${M}_{e}$}}
\newcommand{\nh}{\mbox{${n}_{H}$}}
\newcommand{\rr}{\mbox{${R}_{2}$}}
\newcommand{\rf}{\mbox{${R}_{1}$}}
\newcommand{\vtwo}{\mbox{${V}_{2}$}}
\newcommand{\vout}{\mbox{${V}_{1}$}}
\newcommand{\dshell}{\mbox{${{\rho}_{sh}}$}}
\newcommand{\dwind}{\mbox{${{\rho}_{w}}$}}
\newcommand{\dslow}{\mbox{${{\rho}_{s}}$}}
\newcommand{\dfast}{\mbox{${{\rho}_{f}}$}}
\newcommand{\vfast}{\mbox{${v}_{f}$}}
\newcommand{\vslow}{\mbox{${v}_{s}$}}
\newcommand{\cc}{\mbox{${\rm cm}^{-3}$}}
\newcommand{\apj}{\mbox{ApJ}}
\newcommand{\apjl}{\mbox{ApJL}}
\newcommand{\apjs}{\mbox{ApJS}}
\newcommand{\aj}{\mbox{AJ}}
\newcommand{\araa}{\mbox{ARAA}}
\newcommand{\nat}{\mbox{Nature}}
\newcommand{\aap}{\mbox{AA}}
\newcommand{\gca}{\mbox{GeCoA}}
\newcommand{\pasp}{\mbox{PASP}}
\newcommand{\mnras}{\mbox{MNRAS}}
\newcommand{\apss}{\mbox{ApSS}}

\date{}

\pagerange{\pageref{firstpage}--\pageref{lastpage}} \pubyear{2010}

\maketitle

\label{firstpage}

\begin{abstract}

{

Luminous blue variable (LBV) stars are very massive, luminous,
unstable stars that suffer frequent eruptions. In the last few years,
these stars have been proposed as the direct progenitors of some
core-collapse supernovae (SNe), particularly Type IIn SNe, in conflict
with stellar evolution theory. In this paper we investigate various
scenarios wherein LBV stars have been suggested as the immediate
progenitors of SNe. Many of these suggestions stem from the fact that
the SNe appear to be expanding in a high density medium, which has
been interpreted as resulting from a wind with a high mass-loss
rate. Others arise due to perceived similarities between the SN
characteristics and those of LBVs. Only in the case of SN 2005gl do we
find a valid possibility for an LBV-like progenitor. Other scenarios
encounter various levels of difficulty. The evidence that points to
LBVs as direct core-collapse SNe progenitors is far from
convincing. High mass-loss rates are often deduced by making
assumptions regarding the wind parameters, which are contradicted by
the results themselves. A high density need not necessarily imply a
high wind mass-loss rate: wind shocks sweeping up the surrounding
medium may give a high density shell with a low associated wind
mass-loss rate. High densities may also arise due to wind clumps, or
due to a previous LBV phase before the SN explodes as a Wolf-Rayet
star. Some Type IIn SNe appear to signify more a phase in the life of
a SN than a class of SNe, and may arise from more than one type of
progenitor. A Wolf-Rayet phase that lasts for a few thousand years or
less could be one of the more probable progenitors of Type IIns, and
channels for creating short-lived W-R phases are briefly discussed.

}

\end{abstract}

\begin{keywords}
circumstellar matter; stars: massive; stars: mass-loss; supernovae:
individual: 2005gl; stars: winds, outflows; stars: Wolf$-$Rayet
\end{keywords}

\section{Introduction}
\label{sec:intro}
Core-collapse supernovae (SNe) arise from stars which have a zero-age
main-sequence mass $\ga 8 \msun$, and most likely $\ga 11
\msun$. Surprisingly, the evolution of these stars towards the SN
explosion, and their fate, is not well known after years of intensive
research. Observational programs that set out to unearth the
progenitors of SNe have so far conclusively detected around a dozen,
almost all of which appear to be Type IIP SNe \citep[][and references
  therein]{smartt2009}, with red-supergiant (RSG) progenitors. The
relationship of the remaining types of core-collapse SNe to their
progenitors is inferred but not determined. It had generally been
surmised that stars below about 30 $\msun$ end their lives as RSGs,
whereas those above 30 $\msun$ become Wolf-Rayet (W-R) stars before
they explode as SNe \citep{falk1977, podsi92}. SN 1987A showed that
blue supergiants (BSGs) could also be SN progenitors
\citep{sonnebornetal87}, although it may take a binary companion to
cause them to explode \citep{morris2007, podsiadlowskietal2007}.

Type IIn SNe are relatively recent entrants to the SN classification
scheme \citep{Schlegel1990}. They are characterized by narrow emission
lines (hence the `n' designation) on a broad base, and often (but not
always) show very strong X-ray and radio emission. These
characteristics suggest an interaction with a high-density medium,
resulting in the high X-ray and radio flux. The high density has
furthermore been attributed to a high mass-loss rate
\citep{chugaietal04, galyam2009}. That, combined with other
considerations \citep{kv06, galyam2007, vink2008, smith08,
  trundleetal08, trundleetal2009} has led to suggestions that the
progenitors of Type IIn SNe may be Luminous Blue Variables (LBVs). An
LBV, according to \citet{hd94}, is an ``evolved, very luminous,
unstable hot supergiant which suffers irregular eruptions'' or, in
rare cases ``giant eruptions like $\eta$ Car''. The high mass-loss
rate ($> 10^{-4} \msun {\rm yr}^{-1}$) during the eruptions is mainly
what has prompted the suggestion that LBVs are Type IIn progenitors.

Stellar evolution theorists, on the other hand, have placed LBVs as a
post-main sequence phase, but {\em not} as a final pre-SN phase
\citep{schalleretal92, langer93, Langer1994, Stothers1996,
  maederetal2005, maeder2008}. In conventional models of stellar
evolution theory, the LBV phase may follow a main-sequence O-star
phase or a H-rich WNL phase \citep{crowther08,gh08}, but is {\em
  always} succeeded by a H-poor WN phase or a WC phase. Thus stellar
models require that the LBV star loses its H envelope, becomes a
Wolf-Rayet (W-R) star and then explodes as a SN. Therefore they were
not traditionally considered as progenitors of SNe. The high mass-loss
rates deduced for some SNe such as 1994W \citep{chugaietal04} and
2005gl \citep{galyam2009}, as well as an explosive event that happened
two years before the actual SN explosion in 2006jc \citep{foley2007},
have prompted several authors to question whether LBV's could
indeed be the immediate progenitors of SNe. This has led stellar
evolution experts to explore whether SNe can explode while in the LBV
stage. Several attempts are underway \citep{hirschietal10} but are
limited by the paucity of knowledge regarding LBV and RSG mass-loss
rates, and their metallicity dependence. Although possible scenarios
have been put forward \citep{hirschietal10}, it is fair to say that as
of the writing of this paper, stellar models have so far not succeeded
in having a star in the LBV stage end its life in a SN explosion.

This paper takes a closer look at some of the cases where LBVs have
been proposed as the immediate progenitor of SNe. We explore the
evidence that has been offered for LBV progenitors, and various
reasons why other progenitors have been excluded. Our goal is to
scrutinize various scenarios in detail to determine whether the
evidence requires an LBV as the progenitor, or whether there is room
for alternatives. Irrespective of whether stellar evolution theory
currently allows for SNe to explode in the LBV stage, we wish to
investigate whether the observations require that SNe explode in the
LBV stage.

The plan of this paper is as follows: In \S \ref{sec:LBVprog} we
review proposed scenarios for an LBV star as the progenitor of a
core-collapse SNe. We investigate aspects of the proposed scenario not
considered by the authors, and examine whether the data are compatible
with an LBV progenitor. \S \ref{sec:summary} summarizes our results
and outlines our suggestions, especially where Type IIn SNe are
concerned.

\section[]{Proposed Cases for LBV progenitors of SNe}
\label{sec:LBVprog}
\subsection{SN 2005gl}
\label{sec:2005gl}

This luminous type IIn SN is located in the nearby galaxy NGC
266. Pre-explosion {\it Hubble Space Telescope} (HST) images of the SN
location led \citet{galyam2007} to suggest that a very bright point
source with luminosity $L > 10^6 L_{\odot}$ was the possible
progenitor star of this SN. This identification was confirmed by
\citet[][hereafter GL09]{galyam2009} when they observed that the
putative progenitor had faded away and was no longer visible in {\it
  HST} images. 

Spectra taken 8 days after discovery (between 8 and 33 days after
explosion) show a narrow component to the H$\alpha$ emission with a
velocity of 420 km s$^{-1}$, and an intermediate component with a
velocity of about 1500 km s$^{-1}$ \citep{galyam2009}. No broad
components representing typical SN ejecta velocities in the range of
10$^4$ km s$^{-1}$ is seen. Spectra on day 58 after discovery however
do show a broad component with a 10$^4$ km s$^{-1}$ velocity. The
intermediate component is missing. The narrow component on day 8 is
interpreted by GL09 as the velocity of the unshocked progenitor
wind. The velocity is deemed too high for red supergiants, and too low
for Wolf-Rayet stars, but about right for an LBV star. The dense wind
slows down the SN shock to about 1500 km s$^{-1}$, whose velocity is
reflected in the intermediate component. By day 58, according to GL09,
the shock has crossed the dense wind and entered a region of less
dense wind, and the emission is dominated by fast, unshocked
ejecta. Assuming a steady LBV wind, GL09 determined a mass-loss rate
of about 0.03 $\msun$ yr$^{-1}$, which GL09 contend can only arise
from an LBV star.

The large velocities observed in the day 58 spectrum attest to the
fact that the ejecta have sufficient kinetic energy to expand with
initial velocities $\ga 10^4$ km s$^{-1}$.  Difficulties with this
scenario could be associated with slowing down the fast-moving ejecta
interacting with the ambient medium to about 1500 km s$^{-1}$ in
as little as 8 days, or as much as 33 days depending on the actual
explosion date. We investigate various aspects of this scenario below.

Since the ejecta were expanding freely again on day 58 in this
scenario, we estimate that they would have crossed the dense LBV wind
in 50 -75 days or less (depending on explosion date). By day 8 they
were already at 1500 km s$^{-1}$, and would have slowed down even
more, we may assume that the ejecta had an average velocity about 1250
km s$^{-1}$ for the 50-75 days (depending on explosion date) that it
took to cross the LBV wind medium. Since the velocity of the LBV wind
was about 1/3 that of the SN, it means the wind in this phase
would have lasted for about 3 times more, or about 150-225 days. Given
the mass-loss rate of 0.03 $\msun$ yr$^{-1}$ calculated by GL09, this
means that the total mass ejected in this wind was about 0.0125-0.0185
$\msun$. This is an upper limit, as the wind phase could have
conceivably lasted less than 50-75 days, and the average shock
velocity could have been even lower (note the argument in the previous
paragraph), given that it had slowed down from $ > 10,000$ to 1500 km
s$^{-1}$ in just 8-33 days.

If we make the extreme assumption that all the CS mass is contained in
a thin, dense shell, the properties of the interaction have been shown
to depend on the ratio of the mass of the shell to the mass of the
ejecta \citep{tenorioetal1990, Dwarkadas2005}. A ratio of 1 or higher
is required for the shell to have a significant impact, while a ratio
$<< 1$ would have no impact as the ejecta would just plow through the
shell.  \citet{dwarkadas1997} also showed that the distribution of the
mass, whether it is in a thin shell or spread out, is not as important
as the amount of material. Therefore, in order to decelerate the
ejecta completely, as seems likely, the ejecta mass involved in the
interaction must be smaller than or comparable to 0.02 $\msun$.

The important question therefore is how much ejecta is interacting
with this amount of CSM material. This is difficult to say unless we
can specify the ejecta mass and its distribution, and how the reverse
shock expands into this ejecta, all of which are unknown. However, we
can still make some estimates as to whether the scenario proposed is
consistent with the observations. Following the arguments above,
almost all the kinetic energy of the fraction of the ejecta
interacting with the CSM is lost (the shock velocity decreases to 1/7
the original, therefore the total energy decreases to 1/49 the
original). The total kinetic energy of the amount of ejecta
interacting with the CSM was, at most, 0.5 M$_{ej} {{\rm v}_s}_i^2$,
where ${{\rm v}_s}_i$ is the initial velocity of this material, which
is at least 10,000 km s$^{-1}$. The maximum luminosity that can be
extracted from the radiative shock is 0.5 $\mdot {\rm
  v}_s^3/v_w$. Given the mass-loss rate $\mdot = 0.03 \msun {\rm
  yr}^{-1}$ and the wind velocity $v_w = 420 {\rm km} s^{-1}$ deduced
by GL09, the maximum luminosity that can be extracted immediately
(when the shock velocity is $>$ 10$^4$ km s$^{-1}$), is 4.5 $\times
10^{43}$ ergs s$^{-1}$. However, it is unlikely that the luminosity
will be so high when the shock is moving so fast. On the other hand,
the maximum luminosity on day 8 (33) is 1.5 $\times 10^{41}$ ergs
s$^{-1}$. (Note that \citet{scsff10} find the wind velocity 34 days
after discovery to be closer to 600 km s$^{-1}$, which would reduce
these numbers by about 2/3).  Since as the shock decelerates the
amount of energy released would be expected to increase, we may assume
that the average luminosity for the 8 (33) days was about 1. $\times
10^{42}$ ergs s$^{-1}$. Over 8 days, the total energy released is 6.9
$\times 10^{47}$ ergs, while over 33 days it would be at most 2.85
$\times 10^{48 }$ ergs. The latter is probably an overestimate, as the
shock deceleration would be slower over a larger period. Equating that
to the total energy available for release gives the ejecta mass
interacting with the CSM to be about 1-3 $\times 10^{-3} \msun$, and
possibly up to a factor of 4 smaller. If the initial velocity is
higher this mass increases somewhat, whereas if the wind velocity is
closer to 600 km s$^{-1}$ this decreases some.

The ejecta mass is an order of magnitude smaller than the mass in the
LBV wind it is interacting with. Therefore deceleration of the ejecta
is not an issue. We next check whether this amount of mass in the
unshocked CSM is sufficient to provide the requisite H$\alpha$
luminosity in the narrow line seen by GL09. We assume that the average
velocity of the shock was about 2500 km s$^{-1}$ over the first 8 days
after discovery, or slightly lower if 33 days after
explosion. Therefore, the maximum radius of the shock would be between
about 1.7-7.1 $\times 10^{14}$ cm. This shock would therefore sweep up
a mass of the CSM equal to at most 1.6 $\times 10^{-2} \msun$. The
mass of unshocked CSM remaining is about 2.5 $\times 10^{-3} \msun$ or
larger if the smaller time period is considered. Since the wind
expanded for at most 225 days at 420 km s$^{-1}$, it could have
reached about 8.1e14 cm (somewhat larger if the velocity was closer to
600 km s$^{-1}$). The volume of unshocked wind is about 7.3 $\times
10^{44}$ cm$^{3}$. The luminosity of the H$\alpha$ emission per unit
volume can be approximately written as $\L_{H_{\alpha}} = \alpha n_e^2
h {\nu}_{\alpha}$, where $\alpha = 1-3 \times 10^{-13}$ is the
recombination coefficient. The observed luminosity of 2.8 $\times
10^{39}$ ergs s$^{-1}$ in the narrow line gives the required H number
density of the unshocked wind at that radius to be about 2 $\times
10^9$ cm$^{-3}$. We note that this is consistent with the actual
number density of the unshocked wind just outside the
shock. Furthermore the total mass of H required is smaller than the
mass of unshocked wind. Given the many assumptions that have gone into
this calculation, it appears reasonable that the unshocked wind has
sufficient density to provide the required H$\alpha$ luminosity. Note
that if we take the expansion time as only 8 days then the density
constraint is even easier to satisfy.

One final check to be made is whether the fast moving shock would
become radiative immediately. We consider a shock of 11,000 km
s$^{-1}$ on day 1. The cooling time for such a shock would be around 1
day at a density of 10$^{10}$ cm$^{-3}$, which is expected for the
assumed numbers very close in to the star. Thus it seems likely that
the shock would be a radiative shock, would be able to radiate the
observed amount of energy in the appropriate time period, and that the
ejecta mass would be small enough to be decelerated by the small
amount of LBV material.

We note that the above calculations for the maximum luminosity were
carried out assuming the existence of a steady wind, to be consistent
with the assumptions of GL09, but without any supporting evidence for
this assumption. As we show later in this paper, in many cases a
steady wind is not supported by the observations. However, the above
results could be derived in a different way. A lower limit to the
available luminosity is set by the observed spectrum on day 8,
especially the H$\alpha$ line luminosity. If we assume that the
H$\alpha$ luminosity is about a tenth of the total available
luminosity, the other numbers follow accordingly. Therefore, even if
one removes the steady wind constraint, the computed ejecta mass would
not change much, and would still be lower than the calculated LBV
mass. Removing the steady wind constraint would mean that we would not
be able to easily calibrate the mass-loss rate. However, given the
high density, which is required for the shock to be radiative
(otherwise a broad H$\alpha$ line would be seen on day 8), the
remaining numbers would still be consistent. Also, in order for the
shock wave to become radiative, a high density must exist from day 1
immediately outside the stellar envelope.

{\it Although the calculations are necessarily approximate, and could
  be off by factors of 2-3,} it appears that the interpretation of an
extremely dense medium immediately surrounding the star may be
consistent with most of the observed facts. The fact that the shock
was no longer radiative in the spectrum taken on day 58, given that
the velocity must have been quite close to the initial velocity,
suggests that the density in the wind into which the ejecta are
  now expanding must be lower by more than 2 orders of magnitude, to
make the cooling time at that epoch larger than the flow time. If this
is a wind from an LBV in quiescence, this seems reasonable, as the
mass-loss rate will be at least two orders of magnitude lower. If this
is from a prior phase, say a RSG wind, then since its velocity must be
a factor of 20 or so smaller, this means that the mass-loss rate must
be more than 3 orders of magnitude smaller, or of order 10$^{-5} \msun
{\rm yr}^{-1}$ or lower.

These calculations corroborate the scenario of the SN expanding in a
very high density medium surrounding the star, with a mass-loss rate
of about 0.03 $\msun {\rm yr}^{-1}$ (assuming a steady wind) and a
wind velocity between 420 and 600 km s$^{-1}$. GL09 suggest that these
parameters are typical of LBVs and not found in other stars.  While
this is true, it must be noted that even in LBVs, such high mass-loss
rates are only postulated for LBVs when they are undergoing an
eruption, not when they are in a steady state. The detection of a very
bright source $> 10^6 {\rm L}_{\odot}$ in roughly the V-band is used
by GL09 to support their assertion that this was no ordinary
star. Does this definitely make it an LBV?  Our knowledge of stellar
evolution and the placement of stars on the H-R diagram is incomplete,
and it is unclear if every star with luminosity $> 10^6 {\rm
  L}_{\odot}$ need necessarily be classified as an LBV star.  LBVs are
marked by their variability, which has not, and probably cannot, be
demonstrated in this candidate star. Of concern also is the described
evolutionary scenario that one is forced to accept. In this scenario,
a star transitions into an LBV-like state, undergoes mass-loss with a
mass-loss rate seen only in $\eta$ Car type eruptions for about 6
months or so, and then undergoes core-collapse. Did the star
transition from a quiescent LBV into an eruptive state, and explode
during, or immediately after, that state? Did the star transition from
a different phase (say a red supergiant) to the LBV-like state, making
it an LBV for only 6 months, during which time it underwent an
eruption and then exploded as a SN?  If the explosion of a star as an
LBV is by itself a problem, the explosion of a star immediately after
erupting as an LBV, with an extremely high mass-loss rate, is an even
greater problem for stellar theorists to reconcile.

Therefore, many questions still remain unanswered. Given the available
information, and with the above caveats, we currently rank SN2005gl as
making an acceptable case for an LBV-like progenitor. Any alternative
must have a high density wind (for the given velocity of around 500 km
s$^{-1}$) and a luminosity $> 10^6 {\rm L}_{\odot}$ in the V band, and
satisfy all the constraints computed above.

\subsection{Quasi-Periodic Modulations of Radio Supernovae} 
\citet{ryderetal04} found modulations in the radio light curve of the
Type IIb SN 2001ig, with a periodicity of 150 days (although this is
based on only two peaks, with a probable third). \citet{sckf06}
report similar modulations in the radio lightcurve of SN 2003bg, at
120 and 300 days, with a probably third at 600 days. In this case it
is not clear that they are periodic.  \citet{sckf06} attribute these
to variations in the density of the CS medium. \citet{kv06} suggest
that these density variations may be due to LBV stars undergoing S
Doradus type variations, with the density enhancements specifically
due to the behavior at the bistability limit, which would lead to
density jumps of a factor of 4.

The reason for \citet{kv06} suggesting LBV progenitors was to suggest
a single star scenario rather than more contrived binary star
scenarios. According to \citet{kv06}, the LBV scenario was in line
with both the timescales as well as the amplitudes.  However modelling
of the light curves by \citet{sckf06} has shown that density
enhancements of factors of 1.8, 1.4 and 1.2 are required to fit the
``bumps'' in the light-curve. These are relatively small enhancements,
and it is not clear why an LBV progenitor would need to be invoked to
explain them, as was done by \citet{kv06}. In fact it is somewhat of a
concern as to whether LBV outbursts would give rise to such low-level
density enhancements; on the contrary, they will presumably lead to
much larger changes in density than the 20-40\% required to explain
some of the lightcurve modulations. According to \citet{kv06}
themselves, the bistability jumps lead to a density enhancement of a
factor of about 4. Mass-loss rates and velocities derived around the
bistability jump in \citet{vinketal99} suggest perhaps even larger,
not smaller, density enhancements, and no models suggest 10-20\%
enhancements.  It is therefore unlikely that the density jumps due to
S Doradus variations would match the modelled density enhancements. An
exception would be if both the mass-loss rate and velocity
enhancements were in the same direction, thus leading to a small
enhancement in the density, which depends on their ratio.  It should
also be noted that the models by \citet{sckf06} are approximate, and
not based on hydrodynamical calculations. As they note, the
self-similar solutions they have used are not truly applicable in a
case where the density is increasing, thus leading to some uncertainty
in the calculated results.

Even if bistability jumps were adequate, it is not clear that they
should point only to an LBV.  Although LBVs are the only objects in
which bistability jumps have actually been observed, other massive
stars, besides LBVs, could also experience these bistability
jumps. \citet{vinketal99} have shown that the increase in mass-loss
rate at the bistability jump is mainly due to radiative acceleration
by Fe III, and therefore should occur in hot star atmospheres where
this ion is present. And \citet{vinketal00} have derived mass-loss
rates for all massive O and B stars assuming that they are affected by
the bistability jumps.

An added consideration is that if the enhancements are due to changes
in the wind parameters, then the wind-wind interaction will likely
result in the sweeping up of the outer wind into a dense shell. This
will increase the magnitude of the density enhancement, suggesting
that the actual change in wind parameters may be even smaller, and
making LBV progenitors even less likely.

The timescale calculation is uncertain as it relies on uncertain
velocity determinations, as pointed out by \citet{cs10}. Furthermore,
2003bg was initially classified as a Ic due to lack of hydrogen in the
spectra. It is not known why a H-rich LBV progenitor would give rise
to a Ic spectrum. Perhaps the star transitioned into a W-R star just
before explosion (\S \ref{sec:summary} explores such short W-R
phases). According to the calculations of \citet{cs10} it would have
had to happen on order of a quarter century before explosion. If this
were true, the fast W-R wind would be expected to sweep up the
preceding LBV material into a thin dense shell, which is not
seen. Also, if this were true, an LBV would not be the {\em direct}
progenitor of the SN.

\citet{cs10} suggest that these SNe lie in a category called cIIb, or
compact Type IIb, with radii of a few times 10$^{11}$ cm. They find
that the progenitors of these SNe are less massive and more compact
than LBVs, further discounting LBVs, and perhaps more suggestive of
Wolf-Rayet (W-R) stars. \citet{sckf06} suggested a single W-R star
progenitor with an average mass-loss rate of 3 $\times 10^{-4} \msun
{\rm yr}^{-1}$. However, this mass-loss rate is much higher than the
known mass-loss rate of any single galactic W-R star
\citep{crowther08}. The combination of a high mass-loss rate and a
short W-R progenitor phase may suggest (see also \S \ref{sec:shortWR}
below) that the SN explosion occurred in a binary system
\citep{eit08}. In fact \citet{ryderetal06} claim to have found a
companion star, although \citet{cs10} argue that the star would have a
weak wind and would not create the pinwheel like nebulae that
\citet{ryderetal06} suggest gives the density variations. While
certainly true, this does not preclude the dynamical effects of the
binary companion. It is interesting to speculate whether the stellar
orbit of the companion (rather than its wind), especially if it was
quite eccentric, could affect the mass-loss from the star, leading to
both an enhanced mass-loss rate and a variation in the mass-loss as
the orbital radius changes.

In summary, although we can speculate on the progenitor, we are unable
to pinpoint it. While strong arguments can be made against S Doradus
type variations based on the modeling by \citet{cs10}, there is also
some uncertainty in the models themselves. These issues require
further study and testing via both observations and
modelling. Although we lean towards this scenario not providing
sufficient evidence to suggest LBV supernova progenitors, a final
assessment requires more observations than are currently available,
accompanied by more detailed modelling.

\subsection{Type IIn SNe} 
\label{sec:IIn} 
Many Type IIn SNe show signs of strong CS interaction, indicating that
the SN shock may be encountering a medium of high density
\citep{salamancaetal98, Pastorello2002, salamanca03, cd03,
  chugaietal04, baueretal2008}, which has been interpreted as a sign
of a high wind mass-loss rate. In order to investigate this
assumption, it is useful to understand how the mass-loss rates have
been calculated in several cases. If the medium into which the SN is
expanding is a wind with a constant mass-loss rate $\mdot$ and wind
velocity $v_w$, then the wind density decreases as r$^{-2}$. The mass
swept up by the expanding SN shock up to a radius R is the mass of the
wind upto that radius

\be
\label{eq:msw}
M_{sw} = \mdot R / v_w
\ee

\noi
The total energy in the swept-up material is 

\be
\label{eq:energy}
E = 0.5 M_{sw} v_s^2
\ee

\noi where $v_s$ is the shock velocity. Therefore, the kinetic energy
dissipated per unit time, which is the maximum energy that can be
extracted from the material over time, or the maximum attainable
luminosity, is:

\be
\label{eq:mdot}
L = dE/dt = 0.5 \frac{\mdot}{v_w} v_s^3
\ee

\noi If the luminosity, shock velocity and the wind velocity are
known, then this equation can be used to compute the mass-loss rate,
with perhaps an unknown conversion efficiency factor added. Some form
of this equation has been used to compute the mass-loss rate for
several Type IIn SNe, including SN 1997ab \citep{salamancaetal98}, SN
2005gl, SN 2006gy \citep{scsff10}, SN 2008iy \citep{milleretal10}, and
SN 2006tf \citep{smithetal08}. The important point to note from the
above derivation is that this equation is valid {\em if, and only if,
  the wind mass-loss rate and velocity are constant with time, such
  that its density goes as r$^{-2}$, and the shock velocity is
  constant with time}\footnote{We note here for the sake of
  completeness that, if the shock is radiative, the formula $$ L = 2
  \pi r^2 \rho v^3 $$ can be used to compute the pre-shock density,
  assuming that all the energy is radiated away. This does not require
  a r$^{-2}$ medium, but it does require a radiative shock. However
  substituting $\rho = \frac{\mdot}{4 \pi r^2 v_w}$ to get the same
  equation as above does require assuming a constant mass-loss rate
  and wind velocity. All of the papers discussed here have assumed the
  medium goes as r$^{-2}$ and used equation \ref{eq:mdot} to derive
  the mass-loss rate}. If this is not the case, then equation
\ref{eq:msw} is not valid, and neither are the succeeding equations,
which of course means that equation \ref{eq:mdot} cannot be used. If
it is the answer cannot be relied upon. In general, if the density
profile is not at least as steep as r$^{-2}$, application of these
equations at a given radius will yield a misleadingly high mass-loss
rate.  Unfortunately, in many of the cases that they have been used,
as illustrated below, the values derived contradict the assumptions of
constant mass-loss rate and velocity, and an r$^{-2}$ density profile.

For SN 2006gy, \citet{scsff10} find, by using equation \ref{eq:mdot},
that the value of both $\dot{M}$ and $V_{CSM}$ is varying in time
(their Figure 25). This is contrary to the assumptions inherent in the
equation. Furthermore the SN blast wave is assumed to travel within
this medium with a constant velocity, which is incompatible with the
derived result of a medium with a varying mass-loss rate and velocity
with time. It is almost impossible to construct a model in which the
density parameter $w = \mdot /v_w$ varies non-uniformly by over a
factor of 100 (as the authors find) while still keeping the velocity
of the blast wave roughly constant. Their argument hinges on the
untrustworthy computation of the wind density parameter from equation
[\ref{eq:mdot}]. As they themselves point out, their assumption that
the entire luminosity comes from energy radiated behind a radiative
shock is contrary to their own earlier assumption \citep{Smith2007} of
a shell-shocked diffusion model. If a substantial part of the
luminosity arises from diffusion of energy from a shocked, optically
thick shell, then this would lower their estimates
considerably. Because of the inconsistencies in deriving these
numbers, the total mass of swept-up circumstellar matter (CSM) is also
in doubt, and the high mass-loss rate questionable.

\citet{milleretal10} use equation \ref{eq:mdot} to compute the
mass-loss rate in SN 2008iy. They assume a fixed wind velocity
(although it is measured only at one epoch) and a constant blast wave
velocity. They find that the mass-loss rate increases with time, then
decreases. Given the increase in luminosity over time, they assert
that the parameter $w$ must have increased over 400 days. However, if
the parameter $w$ increases with time, the density profile no longer
goes as r$^{-2}$ over the 400 day rise period, the shock velocity
should certainly not be constant, equation \ref{eq:msw} is not valid,
and neither are the others. Furthermore, although they assume a model
of clumps in a rarefied wind to explain the late-time emission, they
do not take into account any luminosity arising from shocks within the
clumps in their mass-loss rate calculations, which could lower their
estimates significantly. Thus the mass-loss rate is inconsistent with
the assumptions used to obtain it, and therefore unreliable.

\citet{milleretal10} also compute the mass-loss rate by two other
methods: (1) from the X-ray luminosity \citep{ik05} and (2) from the
H$\alpha$ luminosity. Unfortunately, the equation used to compute the
mass-loss rate from the X-ray luminosity makes the same assumption, of
the density of the medium decreasing as r$^{-2}$, which is contravened
by their results. Furthermore the authors use a low X-ray temperature
of 10$^7$K to compute the cooling function at an age of 1.5 years,
which is highly unlikely for a SN at such a young age, and
incompatible with the high blast wave velocity of 5000 km s$^{-1}$
that they have assumed, which would give a post-shock temperature of
few times 10$^8$ K. It is highly unlikely that the spectrum of such a
complex dynamical region could be adequately described by one
temperature, in any case. They have not taken into account the
possibility that the X-ray emission could be arising from both the
forward and reverse shocks. No attempt is made to take the SN
deceleration into account when calculating the radius. Finally, we
note that the numerical coefficient used in this equation is incorrect
due to a missing factor of 4 \citep{immleretal06}.

In their computation of the mass-loss rate from the H$\alpha$
luminosity, it is not clear what velocity \citet{milleretal10} refer
to when using $V_{SN}$. Early in section 4.1 they refer to it as the
velocity of the blast wave overrunning the CSM. This would then make
this the velocity of the forward shock. In a medium as dense as the
one they compute, the shock wave would sweep up the material and
decelerate. The \citet{chevalier1982a} self-similar solution gives the
shock velocity as going as $t^{{\alpha} -1}$ where the deceleration
parameter $\alpha < 1$. The radius would then be given by $R_{SN} =
V_{SN} t_{SN}/\alpha$; the authors have neglected the parameter
$\alpha$. For the method of computing the mass-loss rate from the
H$alpha$ luminosity, they claim that $r_1$ is the ``inner radius''
corresponding to the position of the blast wave, but use the same
formula $r_1 = V_{SN} t_{SN}$, which is incompatible with their
earlier assumption that this is the velocity of the blast wave
overrunning the CSM. It was also assumed that a radius $r_2$ was the
outer radius related to the fast-moving forward shock, and that $r_2
>> r_1$. This is puzzling, because the outer shock is the one
overrunning the CSM, which they earlier referred to as the blast
wave. The assumption that $r_2 >> r_1$ at day 711 would be incorrect
in any model. The \citet{chevalier1982a} solution predicts that the
ratio of the outer to inner radii would be more like 20\% in the
self-similar case. In all cases they have neglected the value of
$\alpha$ in their computations, which would make the radius at each
epoch slightly larger. 

In their scenario to explain the origin of the 400d rise time for this
SN, \citet{milleretal10} favor the clumpy wind scenario. As mentioned,
it is then surprising that they do not take into account the
luminosity from the radiative shocks in the clumps, which would reduce
their mass-loss rate estimates. \citet{milleretal10} also argue
against a wind-bubble scenario on the grounds that after peak the thin
shell is overtaken by the ejecta, and the luminosity is powered by
interaction with the ambient wind. This is not necessarily true. The
shell could be thick, and the peak reached while the shock is still
interacting with the shell. Even when the forward shock exits the
shell and is interacting with the ambient wind, the reverse shock is
still interacting with the shell for some time and could power the
luminosity, and an extremely dense wind may not be needed. Finally,
\citet{milleretal10} say that addition of a wind-bubble component
results in an unnecessary transition from an LBV to a W-R
star. However, this is exactly the kind of scenario we envisage for
many IIns in \S \ref{sec:IIndisc}. Without detailed calculations, a
wind-bubble scenario cannot be excluded. Due to the various
questionable assumptions, their mass-loss estimates are untrustworthy,
and the LBV scenario remains unproven.

In the case of SN 2006tf, the mass-loss rate derived by
\citet{smithetal08}, again using equation \ref{eq:mdot}, exceeds 1
$\msun$ yr$^{-1}$, an astonishingly high rate.  The mass-loss rates
and wind density parameter are found to decrease gradually with time
over 10 years, and decrease by a factor of 10 after that. Note that a
mass-loss rate decreasing outwards, combined with a constant wind
velocity, gives a density profile that decreases faster than
r$^{-2}$. This again means that the results are inconsistent with the
assumptions (constant mass-loss rate and velocity) used to derive
them. \citet{smithetal08} have assumed a shell-shocked diffusion model
to explain the high luminosity, so it seems strange that they would
calculate the mass-loss rate assuming that the entire luminosity
arises from a radiative shock with 100\% energy conversion (which is
inconsistent with the constant shock speed assumed).

The mass-loss rate about 20 years prior to explosion is found to
decrease by a factor of 10, by analysis of the narrow lines. Assuming
that the narrow lines arise from the CSM leads to the high estimate
for the mass-loss rate and total mass. A plausible alternative may be
that the narrow line emission arises from a very small section of the
CSM, such as a denser clumpy component. The ratio of the narrow to
broad line velocity of a factor of 10.5 suggests that the density
contrast between the clumps and the surrounding medium would be about
a factor of 110. The narrow line velocity would not be the velocity of
the CSM, but that of the shock within the clumps. Then the density of
the medium is a factor of 100 lower and does not contribute to the
emission, and the large mass-loss rate, and consequently large mass of
the progenitor, are not necessary. The narrow and broad line
components arise from different entities, and their behavior should
not correlate, as is found for SN 2006tf. This model considerably
diminishes the viability of the LBV hypothesis.

In the above cases, and others in the literature, the high mass-loss
rate was the primary factor in suggesting an LBV progenitor. We have
shown that the high mass-loss rates may be incorrectly
interpreted. This does not mean that the density is not high - a high
density can be achieved independently of a high mass-loss rate. In
fact our main point is that a high density should not be used as a
proxy for a high mass-loss rate, and should not therefore be seen as
indicative of an LBV progenitor. This is elaborated on further in \S
\ref{sec:summary}.

\subsection{Circumstellar Nebulae} 
The similarity in the ring structures of the nebula around the LBV
star HD 168625, the B supergiant Sher 25, the star SBW1, and the
nebula around SN 1987A led \citet{smith07} to argue that the
progenitor of SN 1987A may also have been an LBV. While the first
object is known to be an LBV, the next two have abundances
inconsistent with a RSG phase \citep{smith08}

An argument based purely on the shape of the CS structures is quite
misleading. It is true that the shapes of nebulae around LBV stars and
around SN 1987A are quite similar. However, it should be noted that
the these shapes are also similar to that of several bipolar planetary
nebulae, which form around low mass stars\footnote{see catalog and
  images at \ \\http://www.astro.washington.edu/users/balick/PNIC/} ($
< 8 \msun$). Just as SNe cannot be attributed to be arising from low
mass progenitors on the basis of their circumstellar nebulae, they
cannot be attributed to LBV progenitors purely on the basis of the
shaping of their CS structures. The similarity in the shapes is a
result of the similarity in the shaping mechanism, which is generally
attributed to the interacting winds model and its variations
\citep{kpf78,fm94,dcb96,db98,bf02}, or due to an inherent asymmetry in
the winds themselves \citep{frd98,do02}. The nebula around SN 1987A
has been modelled in a manner similar to the models for PNe
\citep{bl93}. The details of the shaping mechanism may depend on the
parameters, but the basic ideas are similar. This does not imply that
the stars themselves are similar.  Thus the similarity in shapes of
the surrounding medium cannot be argued as a similarity in the
progenitor star.

It could perhaps be argued that while the shaping mechanism may be
effective, in the interacting winds model it does take a change in
wind parameters to bring about this effect. In the case of SN 1987A,
this requirement was fulfilled by the star evolving from the blue to
the red side of the HR diagram and then back to the blue before
exploding. The change in wind parameters, indeed the entire ring
system, has been attributed to the transition from the RSG to the BSG
stage \citep[][and references within]{mccray2007}. If such red-blue
loops were always required for the formation of the ring-like
structures, this would set a timescale requirement which may be
difficult to fulfill. However, we point out that such loops are but
one way of achieving a change in the wind parameters. It has been
reported that RSG and hypergiant stars may also show variability,
related to changes in the wind parameters
\citep{humphreys08,levesque09}, and may even undergo pulsations
\citep{yc10}, as described below. IRC+10420 and VY CMa both show
complex and extensive circumstellar nebulae \citep{humphreys08}. As
discussed below, some stars may experience a brief Wolf-Rayet phase
before explosion. Stars in binary systems can experience mass-loss
episodes due to the pull of the secondary. Massive stars are known to
have weak magnetic fields \citep{puls08}, which can influence the
wind, leading to disk formation and strong shocks, or at the very
least density enhancements towards the magnetic equator
\citep{uo02}. A spherical wind blowing into such a medium may lead to
SN 1987A-like ring structures, as modelled by \citet{bl93}. And
finally, a rotating star may create such shapes due to the latitudinal
asymmetry of the wind from the star itself \citep{do02}. Given our
still incomplete knowledge of stellar mass loss and evolution, it
would be therefore misleading to attribute such shapes purely to LBVs
without concrete evidence.

\subsection{SN 2005gj}

\citet{trundleetal08} find multiple absorption component P-Cygni
profiles of H and He in the spectrum of 2005gj. These profiles have
earlier been noted by \citet{prietoetal07}. \citet{trundleetal08}
interpreted these as indicative of the progenitor's mass-loss
history. Since such profiles have previously been seen only in LBVs,
the natural assumption is to connect SN 2005gj to an LBV
progenitor. The similarities in the wind velocity deduced from the
profiles and those of LBVs was used to further support their argument.

The main problem with an LBV progenitor for SN 2005gj is, that in the
papers describing its discovery and evolution \citep{Aldering2006,
  prietoetal07}, the type attributed to the SN was that of Type
Ia. \citet{prietoetal07} did a cross-comparison with the SNID
database, and the best comparison was overwhelmingly with a Type Ia
spectrum. It is clear that SN 2005gj is not a typical Type Ia, since
it shows Balmer line emission, whose width seems to be increasing with
time. \citet{trundleetal08} appropriately note that the typical S II
and Si II, which solidified the Ia status of the very similar SN
2002ic (since classified as a Ia/IIn), are not as clearly delineated
in SN 2005gj. However, the overall resemblance of the spectrum of SN
2005gj to that of SN 2002ic, between days 26 and 84, is clearly shown
in \citet{prietoetal07}.  The Ia-like features of SN 2002ic are even
accepted by \citet{trundleetal08}. Given this close resemblance
between the two SNe, it seems uncharacteristic that one would be a Ia
and the other have an LBV progenitor. If it is a core-collapse SN, it
may offer an explanation for the absorption components, but the
spectrum becomes more difficult to explain.

The argument therefore boils down to whether SN 2005gj is a Type Ia SN
or not. If it is then it cannot have a massive star LBV progenitor
under any circumstances, and that assumption is not tenable. It is
possible that the multiple absorption components rightly signify
mass-loss, but possibly from a companion star, not the
progenitor. Type Ia's are presumed to occur in a binary system, so the
fact that material from the companion star may surround the progenitor
is not unexpected. This could, and has been used to, explain the
Balmer line emission in a Type Ia \citep{Hamuy2003}.

If it is shown to not be of Type Ia, then the LBV theory becomes more
plausible. But it still needs direct proof that the profiles are due
to the progenitor star as opposed to a companion. Also, it has to be
shown that an LBV was actually the progenitor, as opposed to the fact
that the progenitor star perhaps went through an LBV phase before
losing its H and He envelope and undergoing a brief Wolf-Rayet phase
(see \S \ref{sec:summary}) before explosion. The spectra presented by
\citet{prietoetal07} do show some resemblance to a Type Ic spectrum at
late times. Furthermore, one then has to explain the resemblance of
the spectrum to that of other type Ia SNe, or at least to SN 2002ic,
which would be difficult for a core-collapse event. Therefore, in the
absence of further information, we remain agnostic towards the LBV
interpretation.

\section{Discussion}
\label{sec:summary}

We have discussed in this paper various arguments put forward to
suggest that luminous blue variable stars are progenitors of some
core-collapse SNe.  In many cases the high mass-loss rate that led to
the suggestion of an LBV progenitor was based on the assumption that
the medium into which the SN shock was expanding had a density profile
decreasing as r$^{-2}$, but the results show that the wind mass-loss
parameter is variable with time, implying that the density profile
does not decrease as r$^{-2}$ but is more complicated. Applying an
equation that works for only constant mass-loss rate and velocity
winds to every possible situation, without evaluating whether the
inherent assumptions are satisfied, leads to incorrect
results. Furthermore, unless the density profile is decreasing even
more steeply than r$^{-2}$, application of these equations will
generally yield a mass-loss rate that is higher than the true
mass-loss rate.

The case of SN 2005gl is an exception. The lack of a broad emission
line in the spectrum 8 days after discovery distinguishes it from many
of the other Type IIn. We have examined quantitatively several aspects
of the arguments made by GL09, confirming the basic structure of a
high-density medium surrounding the star. The identification of the
progenitor star as having an exceptionally high luminosity in the V
band (or close to it) is another argument in favor of an LBV
candidate.  However many questions still remain, such as whether an
LBV-like star is the {\em sole} candidate for such a high-luminosity
object, and why an LBV phase could have a high but steady mass-loss
rate, resemblant of an LBV during eruption, for about 6 months, and
then immediately explode as a SN. Given the currently available
information, this could be considered a viable case for an LBV
progenitor, with some unanswered questions remaining

The case for quasi-periodic modulations in the radio data is based on
attributing those variations to S Doradus type instabilities. The
arguments presented in this paper, when combined with those expressed
in \citet{cs10}, seem to contradict the suggestion of an LBV
progenitor here from several different points of view. However more
detailed models and further observations are required for a conclusive
assesment.  The morphological similarity of nebulae around LBV stars
to those around SNe such as SN 1987A in no way implies similarity of
the stars themselves, merely that of the shaping mechanism.

The case of SN 2005gj presents a quandry. It could be considered a
plausible case for an LBV, with reservations, if it can be
conclusively shown that the SN was not of Type Ia, but is not tenable
if the Type Ia interpretation is true. The jury is still out on this
one.

In summary, we have studied several suggested cases for LBV SN
progenitors, and found that in only one case is there significant
evidence to satisfy the interpretation of an LBV-like progenitor.
While the difficulty in doing so may be considerable, it should be
noted that in no case has variability, a hallmark of an LBV, actually
been demonstrated, although it has been alluded to. We emphasize here
that this is not an argument for or against the fact that stars can
explode in the LBV stage, and thereby act as progenitors of SNe. That
is a discussion beyond the scope of this paper, and best left to
stellar evolution theorists who are better equipped to answer that
particular question. What we find here is that {\em in only one
  proposed scenario is a strong case made to invoke an LBV-like object
  as the immediate progenitor}.

Given our incomplete knowledge of stellar evolution, it is important
to determine in the case of SN 2005gl whether other astronomical
objects can have the required high luminosity in the appropriate band,
or whether an LBV is the only candidate. The other arguments, although
consistent and supportive of an LBV interpretation, are not conclusive
on their own. Even one LBV progenitor, if confirmed, does call for a
modification of stellar evolution theory. The challenge for stellar
evolution theorists will then become more difficult, to explain not
only why stars explode in the LBV stage, but whether they can do so
immediately after a high-mass loss rate phase.

At the same time, it appears that an LBV progenitor is not as
pervasive as recent literature tends to suggest. It is clear that
there are several Type IIn for which LBV progenitors, although
suggested, are not necessary. This diversity in IIn progenitors
provides some clues to their understanding, as discussed below.

\subsection{High Density CSM} 
\label{sec:highden} In many individual cases, the assertion 
of an LBV as a SN progenitor is made by noting that the observations
require a high density CSM, and then assuming that the high density
equates to a high mass-loss rate for the progenitor, typically $>
10^{-2} \msun\, {\rm yr}^{-1}$, {\em immediately preceding the SN
  explosion}.  The high mass-loss rate is attributed to an $\eta$
Carinae type LBV explosion. It must be noted however, that even
amongst LBVs, which are not very common in the first place, $\eta$ Car
type explosive ejections are quite rare, with only two known events in
the Galaxy - $\eta$ car itself in the 1840s, and P Cygni. In these
extreme events, the star actually increases in luminosity during the
outburst. In other LBVs with S Doradus type variations, which comprise
the more common variety, the bolometric luminosity of the star does
not vary \citep{pvn08}. It remains to be seen if the frequency of
explosive-type LBV events is high enough to account for all the SNe
that are attributed to it.

Although we have shown that in many cases the high mass-loss rate
argument may not hold, the high X-ray, radio or optical flux suggests
that in many scenarios the CSM density close to the star must be
high. It is therefore prudent to mention here that {\em high density
  does not necessarily imply a high mass-loss rate}. Other mechanisms
can result in a high density medium. A common mechanism is the
interaction of the fast wind from the star with the slower wind from a
previous epoch, forming a wind-blown bubble. The fast wind sweeps the
ambient medium or slower wind into a dense shell, whose density can be
extremely high if the wind shock is radiative (see Figure 1), and the
swept-up mass collapses into a very thin dense shell bordered by a
radiative shock \citep{Weaver1977, Dwarkadas2005}.  The shell density
and thickness depend on the exact wind parameters and interaction
time, but the density can exceed 10$^4 {\rm cm}^{-3}$. The mass in the
dense shell comes from the swept-up material, not the progenitor wind,
and could have been emitted over tens of thousands of years. Depending
on the ambient density, the wind parameters and the time taken for the
interaction, the size of the bubbles and the density enhancements may
vary.  {\em A wind expanding supersonically for a sufficiently long
  time can lead to such dense shells, more-or-less independent of the
  mass-loss rate}. The mass-loss rates cannot then be obtained from
the density using the formulae described in \ref{sec:IIn}.

It is well known that as stars evolve, their wind parameters change,
leading to the formation of large wind bubbles around the star
\citep{gml96, chu03, chu08, fhy03, fhy06, arthur07, dwarkadas07a,
  Dwarkadas2007c}. Moderate changes in the wind parameters lead to the
formation of smaller versions of these bubbles during the last few
thousand years of a star's life. SN 1987A for example is surrounded by
a wind bubble formed about 20,000 years before the SN explosion
\citep{mccray2007}. The progenitor's mass-loss rate must have been
quite low, not exceeding 10$^{-8} \msun {\rm yr}^{-1}$
\citep{chevalier1995, lundqvist1999, Dwarkadas2007b,
  Dwarkadas2007d}. The calculations of \citet{ddb10} infer the
presence of a wind bubble around SN 1996cr, surrounded by a dense
shell at about 0.03 pc.

\begin{figure*}
\includegraphics[angle=0,scale=0.75]{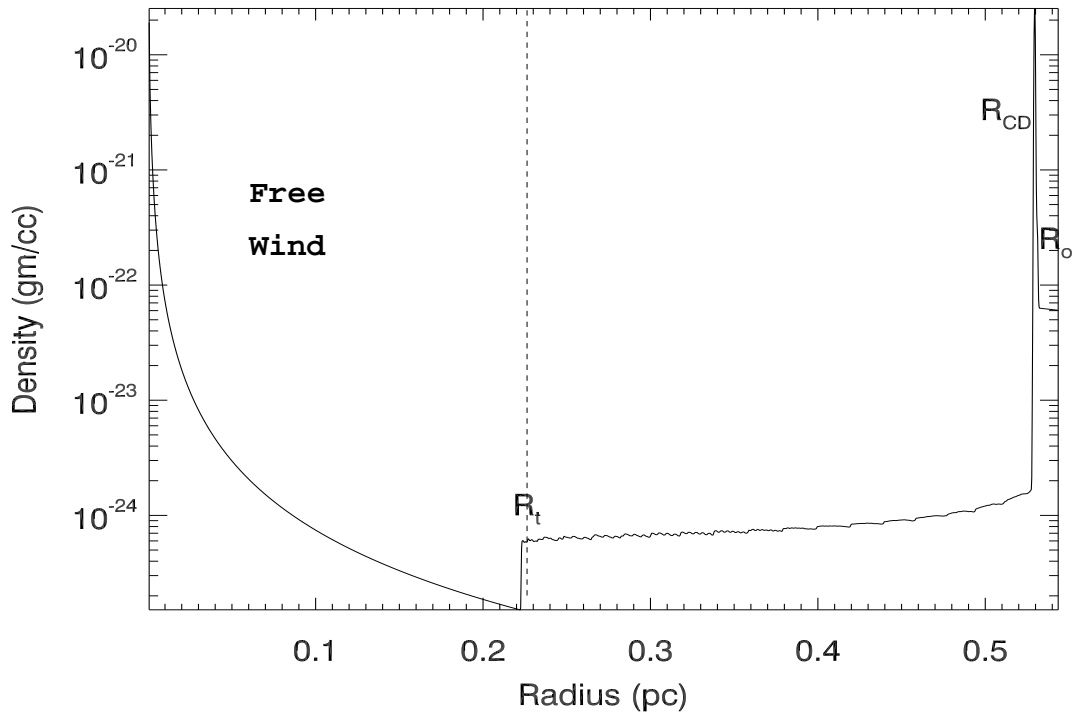}
\caption{The density profile in a wind-blown bubble around a high-mass
  star. The bubble was formed by a Wolf-Rayet wind, with a mass-loss
  rate of 2.5 $\times 10^{-6} \msun {\rm yr}^{-1}$ and a wind velocity
  of 2500 km s$^{-1}$, interacting with a RSG wind from a previous
  epoch. The interaction gives rise to a low-density bubble surrounded
  by a high density shell. This is one way to get a high density in
  the surrounding medium without have an extremely high mass-loss rate
  wind. In this figure, the region marked as ``Free Wind'' denotes the
  freely expanding wind, to which equations \ref{eq:msw} to
  \ref{eq:mdot} can be applied. Application of these equations at
  other radii could result in very large, but incorrect, mass-loss
  rates.
\label{fig:bub}}
\end{figure*}

Figure \ref{fig:bub} shows the structure of such a bubble around a
massive star, taken from a simulation of a Wolf-Rayet wind with
mass-loss rate of 2.5 $\times 10^{-6} \msun {\rm yr}^{-1}$ interacting
with a RSG wind from a previous epoch \citep[see for
  example][]{Dwarkadas2007c}. Note that the density contrast between
the density inside of the bubble and the shell is more than five
orders of magnitude. The wind termination shock ($R_t$), the contact
discontinuity between the two winds ($R_{CD}$) and the outer shock
($R_o$) are marked. Significantly, in this structure, only the region
labelled as free wind (to the left of the wind termination shock R$_t$
and the dotted line) has a density profile that decreases as
r$^{-2}$. Equations \ref{eq:msw} to \ref{eq:mdot} above can {\em only
  be applied to this region to compute the wind parameters}. If these
equations are used to compute the density at other radii, especially
in the dense shell region, it will result in a derived wind parameter
that is extremely high, in a case where there is no freely expanding
wind. This can mislead one into believing that the progenitor is a
high mass-loss rate LBV star. The situation in a star with a varying
mass-loss rate and wind velocity will be even more complicated.

A high density can also be achieved in localized regions due to high
density clumps, which may be formed by hydrodynamical instabilities in
the pre-existing circumstellar medium or in the ejected material. The
interaction of the SN blast wave with a clump leads to a slow-moving
shock expanding into the clump, and a reflected shock back into the
inter-clump medium \citep{kmc94}. The slow-moving shock velocity is
represented in the formation of narrow lines (see \S
\ref{sec:narrowline}). The clumps, although high density and
sufficient to give rise to the narrow lines in the spectrum, make only
a minimal contribution to the total mass of the medium, are not
representative of the overall density of the medium, and cannot be
used to derive the global properties of the medium.

Finally, as current stellar evolution theory suggests for stars that
are more massive than about 40 $\msun$ \citep{hirschietal10}, the
possibility exists that the massive star experienced an LBV phase
before becoming a Wolf-Rayet star. If the Wolf-Rayet stage is short
enough, the dense LBV wind may be close in to the star, and/or may be
swept-up by the faster W-R wind. Either of these possibilities will
give rise to denser material in the surrounding medium, due to an LBV
phase, but does not require the LBV to be the progenitor.  This kind
of scenario was suggested for SN 2001em by \citet{chevalier07b},
following the models of \citet{chugai06}, which showed that a dense
medium around the SN may have formed about 2000 years earlier, with a
mass-loss rate on order $10^{-3} \msun {\rm yr}^{-1}$. The star may
have transitioned from an LBV to a W-R star shortly before explosion.

\subsection{Short Wolf-Rayet Phase}
\label{sec:shortWR}  

In the case of SN 2001em, SN 1996cr, and many others \citep[][and
    references therein; also see \S \ref{sec:IIndisc}]{ddb10}, models
  suggest that the star explodes after a short W-R phase ($< 10^4$
  years). Stellar evolution models, on the other hand, generally show
  that the W-R phase lasts for $>$ 10$^5$ years. Under what conditions
  could we reasonably expect the star to remain in the W-R phase for
  only a few thousand years or less before explosion? It turns out
  that there are several possibilities. (1) One way this could happen
  is if the star has a binary companion. \citet{eit08} have shown that
  stars in a binary system may enter the W-R phase at much lower
  initial mass as compared to single stars. The stars at the very low
  end of the mass range of stars able to become W-R stars will have
  very short W-R lifetimes \citep[see Figures 2 and 3 in][]{eit08}, on
  order tens of thousands of years or less. In principle, the same
  short lifetime would apply to single stars which are at the lower
  end of the mass range where they can become W-R stars, although
  since the cutoff mass to become a W-R is higher, the number of stars
  is much smaller. (2) Binary interaction can result in Case C mass
  transfer which follows the He burning phase. Under certain
  conditions this can lead to explosive common-envelope ejection
  \citep{podsi07,podsietal10}, which leads to the ejection of both the
  hydrogen and helium layers late in the evolution of the star. This
  results in a W-R star with a natural 10$^4$ year or less
  timescale. (3) Another promising channel for the formation of a
  short duration W-R stage was recently shown by the work of
  \citet{yc10}, which followed up on earlier work by
  \citet{hjlb97}. They find that RSGs could have strong pulsation
  driven superwinds. These would cause the RSG to just remove its
  outer layers and become a W-R star in the last hundreds to thousands
  of years of evolution. Such pulsation-driven winds may also result
  in the formation of circumstellar shells around the star, and thus
  may be useful in understanding Type IIns. In fact,
  \citet{fransson02} suggested that Type IIn SNe, and especially SN
  1995N, could arise from RSG progenitors which experience superwinds,
  a scenario that resembles the one outlined above (although they did
  not specifically advocate a W-R progenitor).

Overall it seems that several channels exist which can result in
short-duration Wolf-Rayet phases, and therefore it is reasonable to
propose that some SNe can experience a short W-R phase before
explosion. Many scenarios that require a short W-R phase tend to
indicate stars with initial mass that is lower than the minimum
single-star mass required to become a W-R star. If some fraction of
Type IIn arise from such lower mass W-R stars, then it may indicate
that, contrary to the general assumption, IIns do not arise from very
massive stars but instead from a lower average initial mass
population. A similar result was derived by \citet{aj08}, who found,
from an investigation of the association of the explosion sites of
Type IIn SNe with recent star formation, that the majority of Type
IIns do not arise from the most massive stars, which would be the case
if they all had LBV progenitors.

\subsection{Velocities indicated by Narrow Emission lines}
\label{sec:narrowline}  In many of the 
SNe discussed, and several others, narrow lines seen in the spectrum
are invoked as arising from the unshocked, ionized CSM. In some cases,
the presence of P-Cygni absorption components may indicate the
presence of an outflow. Their velocity is interpreted as the velocity
of the ambient wind, and is considered intermediate between that of a
RSG and a W-R but appropriate for LBVs \citep{trundleetal08}.
However, it is not certain what appropriate velocities are for
LBVs. \citet{trundleetal08} state that LBV velocities lie between
50-300 km s$^{-1}$, \citet{kv06} suggest velocities between 100 and
500 km s$^{-1}$, \citep{scsff10} find 600 km s$^{-1}$ from their
spectrum of SN 2005gl appropriate for an LBV, and $\eta$ Carinae has
reported velocities exceeding 3000 km s$^{-1}$
\citep{smithn08}. Furthermore, such velocities are not unique to
LBVs. Blue supergiants (BSGs) are also known to have velocities of a
few hundred km s$^{-1}$. The circumstellar medium close in to SN 1987A
has been modelled with a BSG wind velocity between 300-600 km s$^{-1}$
\citep{bl93, chevalier1995,Dwarkadas2007b}

As discussed above there are other equally plausible alternatives for
the narrow emission lines. In the cases of SN 1986J \citep{chugai93},
SN 1988z \citep{chugai94}, SN 1978K \citep{cdd95} and SN 1996cr
(Dwarkadas et al.~2010) at least, lines of narrow or intermediate
width are explained as arising from the interaction of the SN shock
with a clumpy medium. The FWHM of these lines denotes the velocity of
the shock driven into the clumps. They are unrelated to the velocity
of the surrounding medium.  There also exists the possibility that the
narrow line spectrum is contaminated by emission from the host
galaxy. We note that clumps could also provide the high density for
much of the emission. Interestingly, in the case of SN 1996cr, both a
dense shell (to provide the enhanced emission) and dense clumps in the
ambient wind (to explain the narrow lines) have been postulated by
\citet{ddb10}.

\subsection{Type IIn SNe} 
\label{sec:IIndisc} The Type IIn SN class is most likely to be 
associated with LBV progenitors. However Type IIn's are not a
homogeneous group, and in fact not all of them show strong radio
emission \citep{vandyketal96} or X-ray emission.  This may suggest
that not all Type IIn are undergoing CS interaction with a high
density medium. The diversity of IIns makes it inappropriate to
classify IIns, or the similarly designated Type Ibn or Icn, as a
separate {\em class} of SNe.  Some are more indicative of a {\em
  phase} in the lifetime of a SN, where it is interacting with high
density material, and which it will outgrow in a matter of years or
decades, no matter what class they fall in or the nature of their
progenitor. A similar point had been made earlier by
\citet{kotaketal04} with regards to the dual classification of SN
2002ic as a Ia/IIn. The fact that Type Ia's sometimes share a dual
classification as Type IIns typifies the inhomogeneity among the
group, and their lack of a common progenitor.

Many IIns earn the `n' designation later on in their lifetime, when
they are presumably interacting with high density material and are
visible. But this merely emphasizes a feature in the late-time
spectrum rather than a physical property of the progenitor. Mainly it
is the initial designation of type which would be useful in
pinpointing a progenitor. This is a small but subtle distinction,
because unlike other SNe classes, it suggests that Type IIn's need not
arise from a single progenitor (if in fact other subtypes do arise
from a single progenitor, which has not been shown). The fact that
IIns could have both high and low initial mass progenitors is
consistent with this statement.

A scenario of a wind bubble blown by a W-R or blue supergiant wind was
suggested for the Type IIn SN 1996cr by \citet{ddb10}, with the
mass-loss rate of the progenitor star being $< 10^{-4} \msun {\rm
  yr}^{-1}$, and in all probability substantially less.  The Type Ibn
SN 2006jc was suggested by \citet{foley2007} to be a Wolf-Rayet star
exploding in a dense He-rich medium formed about two years before the
star exploded. \citet{pastorelloetal08} propose that the progenitor of
the hybrid SN 2005la was a very young Wolf-Rayet (WN-type) star which
experienced mass ejection episodes shortly before core collapse. SN
2003bg was initially classified as a Ic before becoming a IIn, while
SN 2001em transformed from a Ib/c to a IIn, possibly undergoing
substantial mass-loss about 1000-2000 years before explosion
\citep{chugai06}. \citet{stockdaleetal10} indicate that SNe 1996aq and
2004dk, both classified as Type Ic after explosion, are evolving in a
similar manner to SN 1996cr and SN 2001em.

These considerations suggest that a subset of SNe with the `n'
designation may show a wind-bubble structure on parsec or even
sub-parsec scales, due to a change in the wind parameters close to the
end of the star's life.  We propose that one channel for the formation
of Type IIns that show enhanced late-time X-ray and radio luminosity
may be W-R stars with fast winds, where the final W-R phase lasts for
a few thousand years or less (see \S \ref{sec:shortWR}). The W-R wind
sweeps up the surrounding medium, leading to the formation of dense
shells close in to the star.  The interaction of the SN shock wave
with this medium gives rise to many of the notable properties of Type
IIn SNe, including the large X-ray and radio luminosity at late times,
while obviating the need for a high mass-loss rate progenitor (the
mass-loss rates for Galactic W-R stars are $<10^{-4.4} \msun {\rm
  yr}^{-1}$ \citep{crowther08}).

There are some Type IIns that show IIn features almost from the time
of explosion, including SN 2005ip \citep{smithetal09} and SN 1995N
\citep{fransson02}. In these SNe it appears that a high density region
sits just outside the stellar envelope. SN 2005gl also shows an
extremely high density region outside the stellar envelope, but as
seen it does not extend far out radially. These are all indicative of
a wide diversity in this class of objects, and would probably require
a different channel for their formation. SN 2005gl has been associated
with an LBV-like progenitor.  Using {\it Spitzer} spectroscopy,
\citet{foxetal10} have studied SN 2005ip, and conclude that a
progenitor eruption formed a dust shell about 100 years prior to
explosion. They find that a large mass-loss rate $> 10^{-2}$
M$_{\odot}$ yr$^{-1}$ is required to explain this dust shell, and
therefore suggest that it too is consistent with an LBV progenitor. On
the other hand, \citet{smithetal09} found a mass-loss rate 2-3 orders
of magnitude lower. Their optical observations do not require, or
invoke, a dust shell, and they assumed that the mass loss arises in a
steady wind with density decreasing as r$^{-2}$. \citet{smithetal09}
concluded that the progenitor was a RSG star, despite having a wind
velocity of about 120 km s$^{-1}$. It is interesting to speculate
whether these two arguments can perhaps be reconciled by invoking
pulsation-driven superwinds in RSGs \citep{yc10}, which could
potentially give rise to the high mass-loss rate dust shell just prior
to explosion.

Follow-up multi-wavelength observations with increased frequency,
specifically in the radio and X-ray regimes, will undoubtedly reveal
more Type IIn SNe. Regular monitoring may detect more analogs of SN
1996cr, with X-ray and radio emission increasing over year-long
periods of time. Such observations, accompanied by detailed
theoretical investigations including hydrodynamic modelling and
multi-wavelength emission computations, are urgently needed, if we are
to determine the nature and progenitors of this interesting category
of objects.

\section*{Acknowledgments}

I gratefully acknowledge discussions with Alex Heger and Georges
Meynet, who as usual patiently answered my queries on stellar
evolution, and directed me to various literature regarding the
formation of Wolf-Rayet stars with a short lifetime.  I thank Roger
Chevalier for reading an earlier version of this manuscript and
providing suggestions that helped to improve it. I also wish to
acknowledge a very comprehensive report by the anonymous referee,
which considerably helped to improve and strengthen the arguments in
this paper. VVD's research is supported by grants TM9-0001X,
TM9-0004X, GO9-0086B, and GO0-11095B, provided by the National
Aeronautics and Space Administration through Chandra Awards issued by
the {\it Chandra} X-ray Observatory Center, which is operated by the
Smithsonian Astrophysical Observatory for and on behalf of the
National Aeronautics Space Administration under contract NAS8-03060. I
thank my collaborators in our research on SN 1996cr, Dan Dewey and
Franz Bauer, which led me to look much more closely at LBV
progenitors.

\bibliographystyle{mn2e}
\bibliography{paper}

%% This figure uses \includegraphics to scale and rotate the still frame
%% for an mpeg animation.

\bsp

\label{lastpage}

\end{document}